\documentclass[onecolumn,
]
{revtex4}
\usepackage{amssymb}
\usepackage{epsfig}
\newcommand{\beq}{\begin{eqnarray}}
\newcommand{\eeq}{\end{eqnarray}}
\newcommand{\me}{\frac{1}{2}}
\begin{document}

\title{Perturbative self-interacting scalar field theory: a differential equation approach.}

\author{R. da Rocha}
\email{roldao@ifi.unicamp.br}
\affiliation{Instituto de F\'{\i}sica Gleb Wataghin, Universidade Estadual de Campinas,
CP 6165, 13083-970\\ Campinas, SP, Brazil.}

\author{E. Capelas de Oliveira}
\email{capelas@ime.unicamp.br}
\affiliation{Departamento de Matem\'atica Aplicada, IMECC, Unicamp, CP6065, 13083-859, Campinas, SP, Brazil.}

\author{C. H. Coimbra-Ara\'ujo}
\email{carlosc@ifi.unicamp.br}
\affiliation{Instituto de F\'{\i}sica Gleb Wataghin, Universidade Estadual de Campinas,
CP 6165, 13083-970\\ Campinas, SP, Brazil.}


\begin{abstract}
We investigate the partition function related to a   $\phi^4$-scalar field theory on a $n$-dimensional Minkowski spacetime, which is shown
to be a self-interacting scalar field theory 
at least in 4-dimensional Minkowski spacetime. After revisiting the analytical calculation of the perturbative expansion coefficients and also 
the approximate values for suitable limits using Stirling's formul\ae, we investigate a spherically symmetric scalar field in a $n$-dimensional 
Minkowski spacetime. For the first perturbative expansion coefficient it is shown how it can be derived a 
modified Bessel equation (MBE), which solutions are investigated in one, four, and eleven-dimensional Minkowski spacetime.
The solutions of MBE are the first expansion coefficient of the series associated with the partition function of $\phi^4$-scalar field theory. All results are shown graphically.  
\end{abstract}

\maketitle

\section{Introduction}

The results presented in this article are based on Witten's proposed questions, solved in \cite{qfts}, by P. Deligne, 
D. Freed, L. Jeffrey, and S. Wu, concerning perturbative $\phi^4$-scalar field theory (PSFT). We extend their solutions
and show that the first perturbative coefficient can be obtained from  modified Bessel functions of first kind.
This article is organized as follows:  In Sec. (\ref{s1}) the partition functional is exhibited in the context 
of a self-interacting $\phi^4$-scalar field theory and a perturbative solution for the partition function
is obtained, in the light of the solutions obtained by Deligne, Freed, Jeffrey, and Wu \cite{qfts}.
 In Sec. (\ref{s2}) the method presented is shown to hold for a spherically symmetric  scalar field in $n$-dimensional 
Minkowski spacetime. The perturbative expansion coefficients are derived, and the first coefficient is
shown to constrain the scalar field in a MBE, which solutions are used to obtain the first 
perturbative expansion coefficient associated with the partition function of PSFT, for the particular cases
of 1-, 4-, and  11-dimensional spacetime. Numerical integration permits us to accomplish the expansion coefficient 
for 2- and 4-dimensional spacetime. In Appendix we plot the main results.

\section{Perturbative $\phi^4$-scalar field theory}
\label{s1}
We begin with the general form of the scalar Lagrangian density
\begin{equation}\label{1}
{\mathcal L} = \me\partial_\mu\phi\partial^\mu\phi - V(\phi),
\end{equation}\noindent where $\phi=\phi(x)$ is a Hermitian scalar field and  $V = V(\phi(x))$ denotes the scalar potential.
For instance, in a self-interacting 4-dimensional scalar field theory it is well-known that \cite{ramond}
\begin{equation}\label{2}
V(\phi) = \me m^2\phi^2 + \frac{1}{4!}\lambda\phi^4,\end{equation}\noindent  where $\lambda$ is the self-coupling constant and 
$m^2$ denotes the mass parameter.  Euler-Lagrange equation of motion gives, from eq.(\ref{1}),
\begin{equation}
\partial_\mu\partial^\mu\phi = -\frac{\partial V(\phi)}{\partial\phi}.
\end{equation}\noindent 

The partition function associated with eqs.(\ref{1},\ref{2}), can be written as
\begin{equation}\label{22}
Z(\lambda) = \int_{-\infty}^{\infty} \exp\left(-\me\phi^2 - \frac{1}{4!}\lambda\phi^4\right)\,d\phi.
\end{equation}\noindent As we want to investigate the perturbative character of the scalar field theory given by the 
Lagrangian density in eq.(\ref{1}) with scalar potential given by eq.(\ref{2}), 
it is now possible to expand the partition function $Z(\lambda)$ in terms of
a series as 
\begin{equation}\label{4}
Z(\lambda) = \sum_{k=0}^\infty c_k\lambda^k,\quad c_k\in\mathbb{R}.
\end{equation}\noindent
Using Taylor's formul\ae\, it can be shown that \cite{qfts}
\begin{equation}\label{3}
c_k = \frac{(-1)^k}{(4!)^k k!} \int_{-\infty}^\infty \exp\left(-\me\phi^2 - \frac{1}{4!}\phi^{4k}\right)\,d\phi.
\end{equation}\noindent Now define 
\beq
f(A) &=& \int_{-\infty}^\infty \exp\left(-\me\phi^2 + A\phi\right)\,d\phi \nonumber\\
&=& \int_{-\infty}^\infty \exp\left(-\me(\phi + A)^2\right) \exp\left(\me A^2\right)\,d\phi\nonumber\\
&=& \sqrt{2\pi}\exp\left(\me A^2\right).
\eeq\noindent It is clear that the integral involving the coefficient $c_k$ in eq.(\ref{3}) can be written as 
\begin{equation}
\frac{\partial^{(4k)} f(A)}{\partial A^{(4k)}}\arrowvert_{A=0} = \sqrt{2\pi}\frac{(4k)!}{(2k!) 4^k},
\end{equation}\noindent from which it follows that \cite{qfts}
\begin{equation}
c_k = \frac{(-1)^k(4k)!\sqrt{2\pi}}{k! (24)^{k}(2k)! 4^k} = (-1)^k\frac{\sqrt{2\pi}}{(24)^k} (4k-1)!!
\end{equation}\noindent Then, perturbatively for large $k$ the expansion in eq.(\ref{4}) gives for the coefficient $c_k$ the expression
\begin{equation}
c_k \approx (-1)^k \sqrt{2} k^{-1/2}\left(\frac{2k}{3}\right)^k e^{-k}
\end{equation}\noindent where Stirling's formul\ae$\;$ $n!\approx \sqrt{2\pi n}\,n^n\,e^{-n}$ is used \cite{mat,qfts}.

\section{$n$-dimensional spherically symmetric path integral: perturbative
method}\label{s2}
Considering a spherically symmetric scalar field $\phi = \phi(r)$ in a $n$-dimensional spacetime, 
where $r$ denotes 
the radial coordinate, the Lagrangian density is given by 
\begin{equation}
{\mathcal L} = \me \|d\phi\|^2 + \me m^2\phi^2 + \frac{1}{4!}\alpha^{4-n}\lambda\phi^4,
\end{equation}\noindent where $m$ denotes the mass associated with the scalar field and $\alpha$ is an arbitrary mass \emph{constant}
parameter introduced in order to leave the  self-coupling constant $\alpha$ dimensioless in $n$-dimensions. The term $d\phi$ denotes
the differential operator acting on the scalar field $\phi$ as $d\phi = \displaystyle\frac{\partial\phi}{\partial r} dr$.

The standard path integral representation of a theory defined by an action $S = \int {\mathcal L}$ is obtained by defining the quantity 
$Z$, the partition function, as 
\begin{equation}\label{11}
Z = \int  \exp (iS)\, {\mathcal D}\phi\end{equation}\noindent where the $\int {\mathcal D}\phi$  expression is  just a shorthand  for the product 
\begin{equation}
\int {\mathcal D}\phi = \prod_{x} \int d\phi(x).
\end{equation}\noindent This product is taken over all infinite spacetime points $x$ in the volume of the system being described. 
We immediately spot similarities between the path integral in eq.(\ref{11}) and the partition function of statistical mechanics $Z = {\rm tr}\, [\exp(-S)]$ 
\cite{bel}. 
The above integral over all field configurations is precisely analogous to a trace over all degrees of freedom of a system. 
However, the imaginary exponential factor in the path integral does not lend itself to a probabilistic interpretation 
in the same way that with the partition function. 
A standard trick used for formulating theories on a lattice is employed, 
which is to move into imaginary or Euclidean time \cite{bel}. The Euclidean spacetime metric tensor 
is given by $g_{ij} = \delta_{ij}$, where $\delta_{ij}$ denotes the Kronecker tensor.
 The path integral 
is written as 
\begin{equation}
 Z = \int  \exp(-S) \, {\mathcal D}\phi
\end{equation}\noindent  where
\begin{equation}
S = \int \left(\me \partial_\mu\phi \partial^\mu\phi + \me m^2\phi^2 + \frac{1}{4!}\alpha^{4-n}\lambda \phi^4\right) \;dt\wedge dx_1\wedge\cdots\wedge dx_{n-1} \end{equation}\noindent
 with $\partial_\mu\partial^\mu = \displaystyle\sum_{i=1}^{n-1}\frac{\partial^2}{\partial x_i^2} + \frac{\partial^2}{\partial t^2}$. The exponential function in the path integral 
is now free of imaginary factors and can be interpreted as a probability distribution, thereby making the connection with the standard partition
 function from statistical mechanics. The Euclidean formulation is crucial for the operation of lattice Monte Carlo investigations \cite{arn}.  Hereon
\begin{equation}
d\eta = dt\wedge dx_1\wedge\cdots\wedge dx_{n-1}\end{equation}\noindent  denotes 
the $n$-dimensional volume element in a given local coordinate chart.

Proceeding as in Sec. (\ref{s1}), by expanding the partition function as in eq.(\ref{4}) 
it follows that
\beq
c_k &=& \frac{(-1)^k}{(4!)^k k!} \int \left[\exp\left[-\me\int(\|d\phi\|^2 + m^2\phi^2) d\eta\right]  \left(\int\alpha^{n-4}\phi^4d\eta\right)^k \right]{\mathcal D}\phi\label{6}\\
&=& \frac{(-1)^k}{(4!)^k k!} \int \exp\left[-\me\int (\|d\phi\|^2 + m^2\phi^2)d\eta + k\ln \left(\int \alpha^{n-4}\phi^4\,d\eta\right)\right]\; {\mathcal D}\phi.\label{7}
\eeq\noindent In order to find the exponent critical points in eq.(\ref{7}), which is equivalent to have knowledge of the large $k$ behavior of the coeffiecients $c_k$ 
\cite{qfts},
one takes the derivative with respect to the $r$ coordinate in both terms of the
 exponent $\displaystyle{-\me\int (\|d\phi^2\| + m^2\phi^2)d\eta + k\ln \left(\int \alpha^{n-4}\phi^4\,d\eta\right)}$, yielding:
\begin{equation}\label{8}
\int \left(-\Delta\phi - m^2\phi\right)\frac{d\phi}{dr}\,d\eta + \frac{4k\int \phi^3\frac{d\phi}{dr}\,d\eta}{\int\phi^4\,d\eta}
 \end{equation}\noindent where $\Delta$ denotes the Laplacian operator. At the critical points of the exponent, the integrand in eq.(\ref{8}) equals zero, and then
\begin{equation}\label{19}
(\Delta + m^2)\phi = \frac{4k\phi^3}{\int\phi^4d\eta}.\end{equation}
\noindent As $\phi=\phi(r)$ is a radial scalar field, the Laplacian is given by 
\begin{equation}
\Delta\phi = -\frac{1}{r^{n-1}}\frac{d}{dr}\left(r^{n-1}\frac{d\phi}{dr}\right),\end{equation}\noindent 
and eq.(\ref{19}) can be written as
\begin{equation}
\frac{d^2\phi}{dr^2} + \frac{n-1}{r} \frac{d\phi}{dr} - m^2\phi + \frac{k\phi^3}{\pi^2\int_0^\infty r^3\phi^4(r)dr} = 0
\end{equation}\noindent which can be led to
\begin{equation}\label{9}
\frac{d^2\phi}{dr^2} + \frac{n-1}{r} \frac{d\phi}{dr} - m^2\phi + k\phi^3 = 0
\end{equation}\noindent if the rescaling $\phi\mapsto \left(\pi^2\int_0^\infty r^3\phi^4(r)dr\right)^{-2}\phi$ is imposed \cite{qfts}. This rescaling 
must be finite, and so the condition $\displaystyle\lim_{r\rightarrow\infty}\phi(r) = 0$ must holds. Besides, $\displaystyle\lim_{r\rightarrow 0}\frac{d\phi}{dr} = 0$ in order that
the derivative $d\phi/dr$ to make sense in $r = 0$ \cite{qfts}.
Denoting hereon 
\begin{equation}\label{10}
V(\phi) = \frac{k}{4}\phi^4 - \frac{m^2}{2}\phi^2,\end{equation}\noindent for $n=1$ eq.(\ref{9}) is Newton's second law associated with the potential $V(\phi)$ given by
eq.(\ref{10}):
\begin{equation}
\frac{d^2\phi}{dr^2} + \frac{dV(\phi(r))}{d\phi} = 0,
\end{equation}\noindent For $n>1$ we have 
\begin{equation}
\frac{d^2\phi}{dr^2} +  \frac{dV(\phi(r))}{d\phi} + \frac{n-1}{r}\frac{d\phi}{dr} = 0.
\end{equation}
\noindent When $k=0$, eq.(\ref{9}) can be written as 
\begin{equation}\label{20}
\frac{d^2\phi}{dr^2} + \frac{n-1}{r} \frac{d\phi}{dr} - m^2\phi = 0
\end{equation}\noindent
which can be exactly solved if we first multiply eq.(\ref{20}) by $r^2$, yielding
\begin{equation}\label{20}
r^2\frac{d^2\phi}{dr^2} + r(n-1) \frac{d\phi}{dr} - m^2r^2\phi = 0.
\end{equation}\noindent Now, let us consider $\phi(r) = r^\beta \xi(r)$, and on substituting this \emph{ansatz}
in the equation above, which seems to be the best adapted to our investigation, it follows that
\begin{equation}\label{31}
r^2\frac{d^2\xi(r)}{dr^2} + (2\beta + n-1) r\frac{d\xi(r)}{dr} + \left[\beta(\beta-1) + (n-1)\beta\right]\xi(r) - m^2r^2\xi(r) = 0.
\end{equation}\noindent We now are left with the task of determining a suitable choice of the parameter $\beta$ in eq.(\ref{31}) 
in such a way that the coefficient of the term $\displaystyle{r\frac{d\xi(r)}{dr}}$ to be 1, i.e., we choose 
$\beta$ in order that eq.(\ref{31}) can be led to a MBE. With this choice, 
it can be immediately shown that $\beta = 1 - \displaystyle\frac{n}{2}$.
Eq.(\ref{31}) is then expressed
as
\begin{equation}
r^2\frac{d^2\xi(r)}{dr^2} + r\frac{d\xi(r)}{dr} -\left[\left(1-\frac{n}{2}\right)^2 + m^2r^2\right]\xi(r) = 0.
\end{equation}\noindent This is the MBE, which has as solutions 
the modified Bessel functions of the first kind $I_{1-\frac{n}{2}}(mr)$ and of the second kind $K_{1-\frac{n}{2}}(mr)$ \cite{abra}. 
Finally the self-interacting radial scalar field is given, for $k=0$, by 
\begin{equation}
\phi(r) = r^{1-\frac{n}{2}}\;I_{1-\frac{n}{2}}(mr).
\end{equation}\noindent It can be shown that the solution above is regular at the origin, at least for $n$ even \cite{mat}.
The graphics in Appendix show the behavior of such function for each $n$. We use unitary field mass, and assume $m=1$ without loss
of generality.

From eq.(\ref{3}) it follows that the first partition function perturbative expansion coefficient $c_0$ in eq.(\ref{4}) is given by  
\begin{equation}\label{37}
c_0 =  \lim_{\epsilon\rightarrow 0} \int_{\epsilon}^\infty \exp\left(-\me r^{2-n}\;I_{1-\frac{n}{2}}^2(mr)\right)\, \frac{d I_{1-\frac{n}{2}}}{dr}\,dr.
\end{equation}\noindent 
For $n=1$ we have
\begin{equation}
c_0 = \sqrt{\frac{2}{\pi m}}\lim_{\epsilon\rightarrow 0}\int_\epsilon^\infty \frac{1}{r}\exp\left[-\sqrt{\frac{r}{2\pi m}} \sinh(mr)\right]\,\left(-\frac{\sinh(mr)}{2r} + m\cosh(mr)\right)\,dr
\end{equation}\noindent 
For $n=4$, consisting of the usual Minkowski spacetime $\mathbb{R}^{1,3}$, it follows that $c_0$ is given by 
\begin{equation}
c_0 = \int_0^\infty \exp\left[-\frac{m^3}{16}\left(\sum_{j=0}^\infty \frac{(mr)^{2j}}{4^j j! (2+j)!}\right)^2\; \left(\sum_{p=0}^\infty \frac{(2p+1)(mr)^{2p}}{4^p p! (2+p)!}\right) \right] dr, 
\end{equation}\noindent and numerical integration gives us
\begin{equation}
c_0 = 3.85378.
\end{equation}
For $n=2$, it follows from eq.(\ref{37}) that
\begin{equation}
c_0 = 0.39769.
\end{equation}

Finally, for $n=11$, where now the scalar field is a 11-dimensional  Minkowski spacetime-valued function,
it follows that
\beq
c_0 &=& \sqrt{\frac{1}{2\pi m^9}}\lim_{\epsilon\rightarrow 0}\int_\epsilon^\infty \frac{dr}{r}\exp\left[\sqrt{\frac{2}{\pi (mr)^9}}\,(m^4r^4 + 45m^3r^2 + 105)\sinh(mr) - mr(10m^2r^2 + 105)\cosh(mr)\right]\nonumber\\
&&\times \{\sinh(mr)(-14m^4r^5 - 9m^4r^4 - 30m^2r^3 - 405m^2r^2 - 945) + mr\cosh(mr)(2m^4r^5 + 30m^2r^3 + 90m^2r^2 + 945)\}\nonumber
\eeq\noindent This formul\ae \, can be useful in investigating the KK gravitational waves propagation modes in brane-world scenario \cite{Gregory, nosso}.

\section{Concluding Remarks}
The first perturbative expansion coefficient of the partition function associated 
with a self-interacting $n$-dimensional Minkowski spacetime-valued scalar field is obtained from the exact solution of MBE,
and the particular cases of one, four, and eleven-dimensional spacetimes are obtained in terms of modified Bessel funcions of first kind.
The 11-dimensional Minkowski spacetime case is shown to be very useful when we compactify 
it to an AdS$_5 \times S^7$ spacetime, which will be discussed  and investigated in the context 
of Kaluza-Klein modes of gravitational waves propagation \cite{nosso}.

Anomalous dimensions associated with the operators of type $\phi^d$, $d\in\mathbb{N}$, 
in a 1- and 2-loop approximations \cite{qfts, ramond} are to be investigated in the light of the
other coefficients $c_k$, $k>0$ in a forthcoming paper.

\section*{Appendix}

\begin{figure}
\centering
\includegraphics[width=8.7cm]{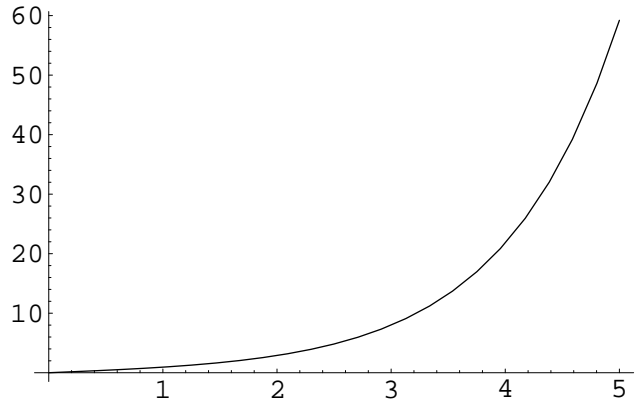}
  \caption{\small $\phi(r)\times r$ evaluated for $n=1$.}
\label{n1}
\end{figure}

\begin{figure}
\centering
\includegraphics[width=8.7cm]{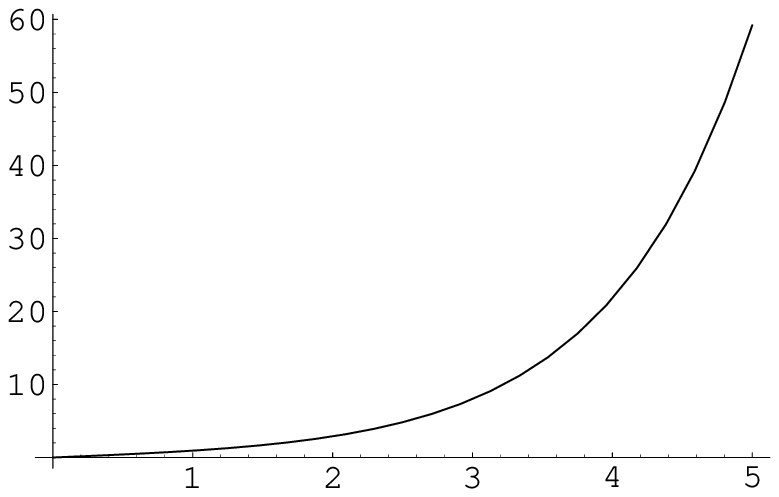}
  \caption{\small $\phi(r)\times r$ evaluated for $n=2$.}
\label{n1}
\end{figure}

\begin{figure}
\centering
\includegraphics[width=8.7cm]{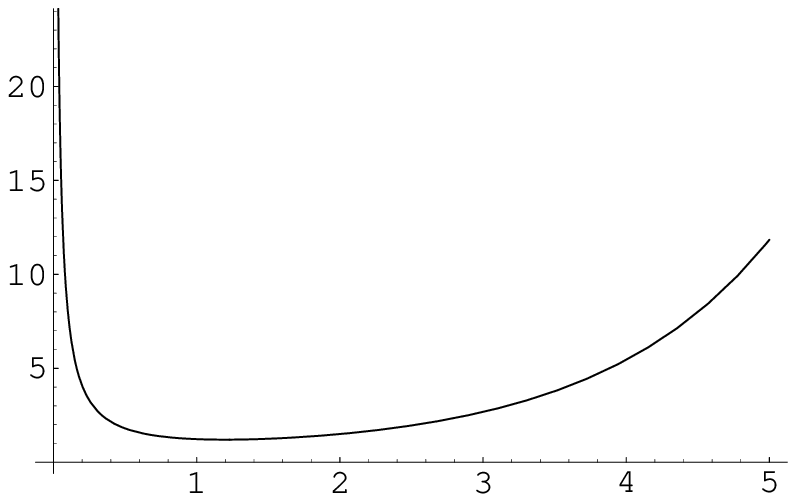}
  \caption{\small $\phi(r)\times r$ evaluated for $n=3$.}
\label{n1}
\end{figure}

\begin{figure}
\centering
\includegraphics[width=8.7cm]{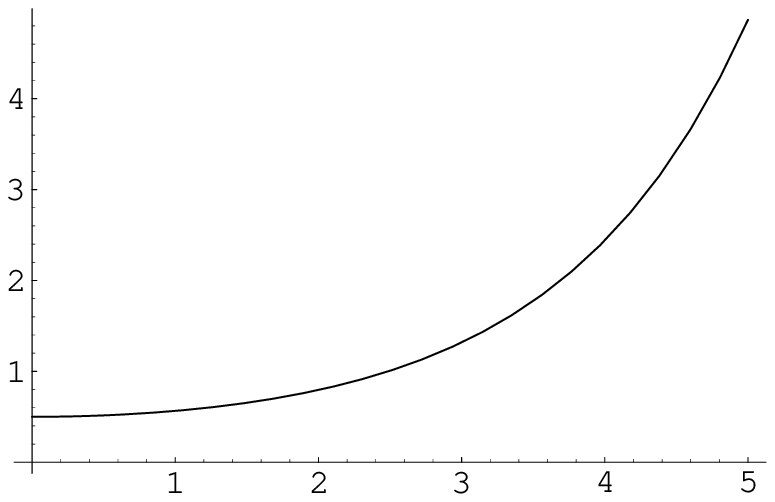}
  \caption{\small $\phi(r)\times r$ evaluated for $n=4$.}
\label{n1}
\end{figure}

\begin{figure}
\centering
\includegraphics[width=8.7cm]{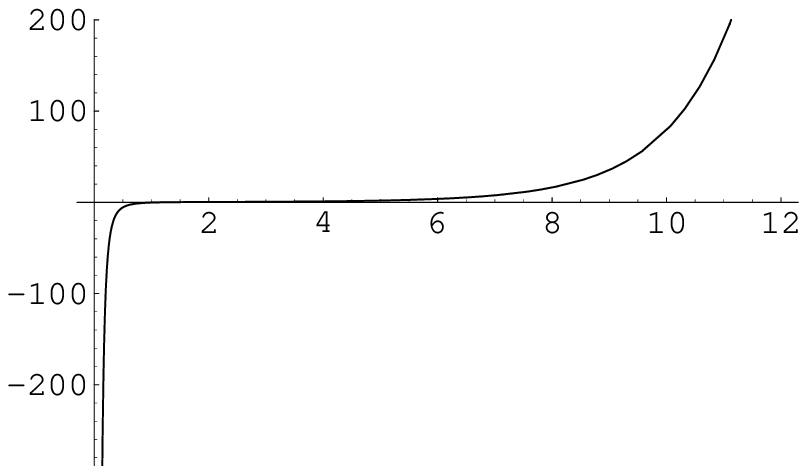}
  \caption{\small $\phi(r)\times r$ evaluated for $n=5$.}
\label{n1}
\end{figure}

\begin{figure}
\centering
\includegraphics[width=8.7cm]{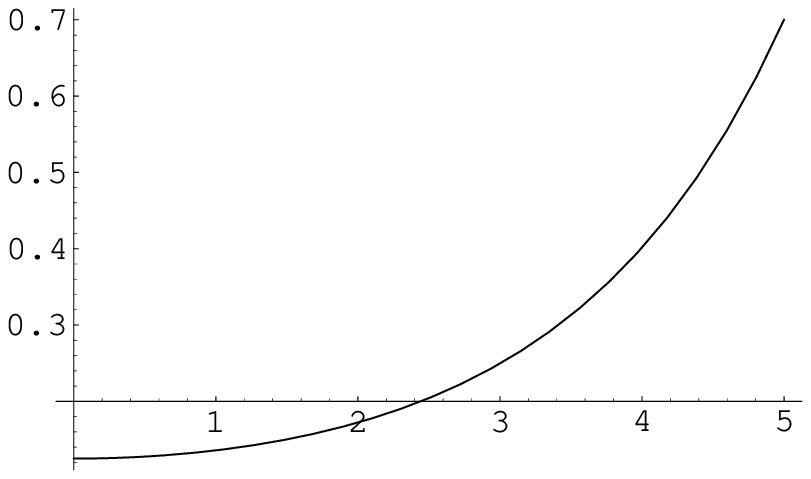}
  \caption{\small $\phi(r)\times r$ evaluated for $n=6$.}
\label{n1}
\end{figure}

\begin{figure}
\centering
\includegraphics[width=8.7cm]{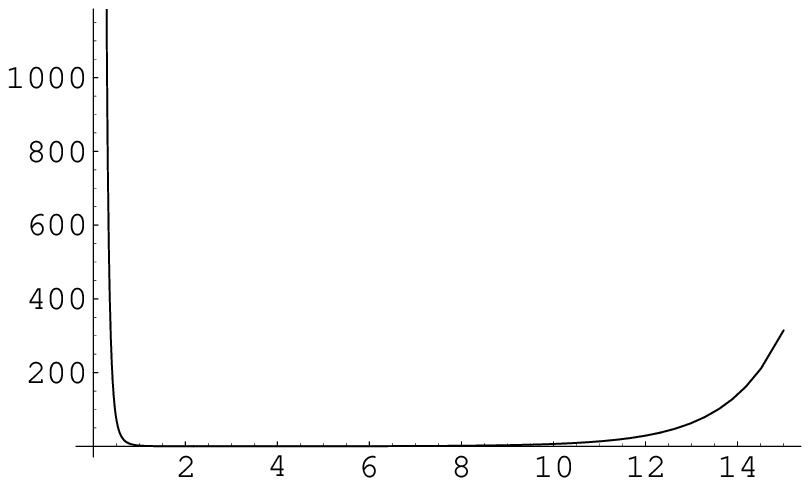}
  \caption{\small $\phi(r)\times r$ evaluated for $n=7$.}
\label{n1}
\end{figure}

\begin{figure}
\centering
\includegraphics[width=8.7cm]{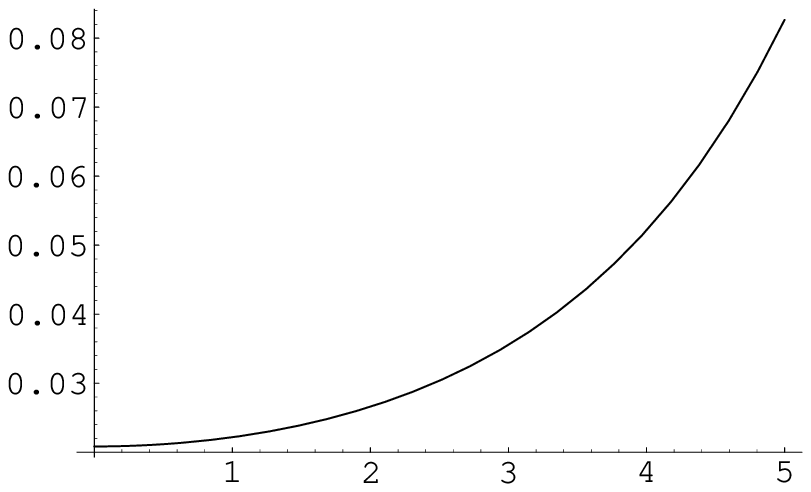}
  \caption{\small $\phi(r)\times r$ evaluated for $n=8$.}
\label{n1}
\end{figure}

\begin{figure}
\centering
\includegraphics[width=8.7cm]{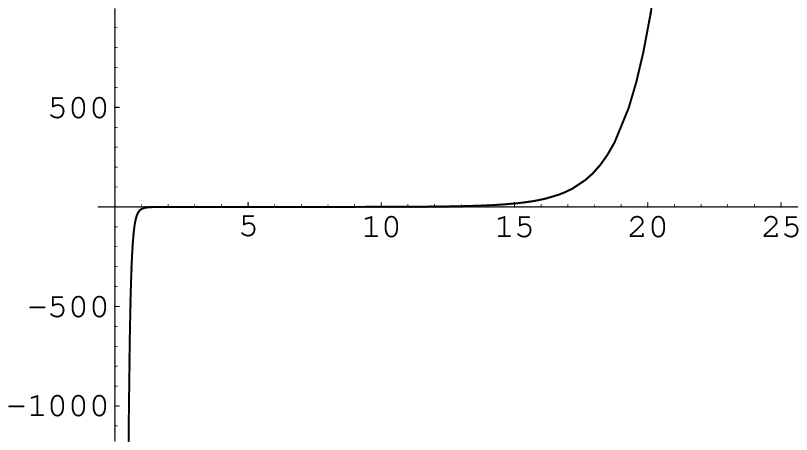}
  \caption{\small $\phi(r)\times r$ evaluated for $n=9$.}
\label{n1}
\end{figure}

\begin{figure}
\centering
\includegraphics[width=8.7cm]{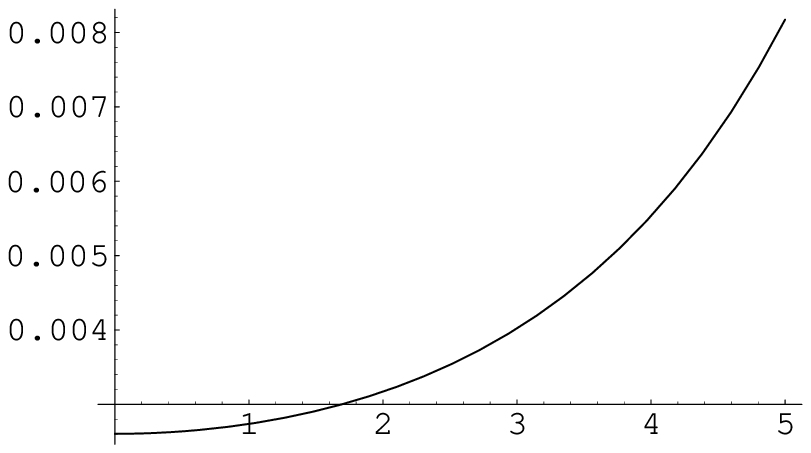}
  \caption{\small $\phi(r)\times r$ evaluated for $n=10$.}
\label{n1}
\end{figure}

\begin{figure}
\centering
\includegraphics[width=8.7cm]{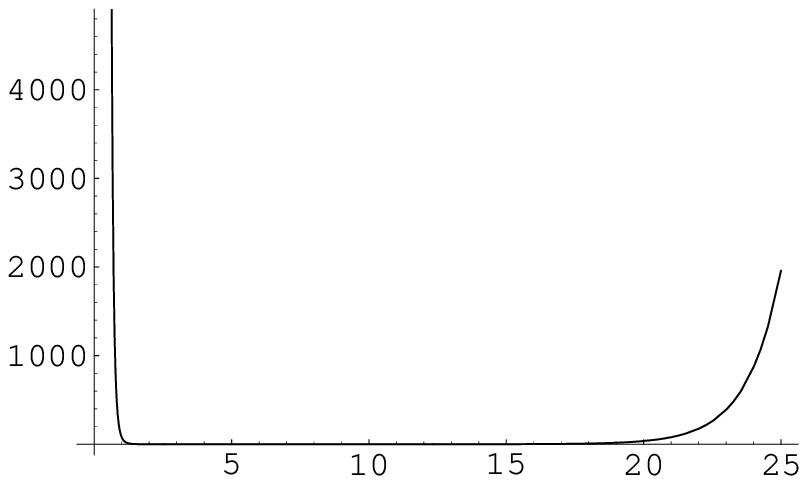}
  \caption{\small $\phi(r)\times r$ evaluated for $n=11$.}
\label{n1}
\end{figure}

\begin{figure}
\centering
\includegraphics[width=8.7cm]{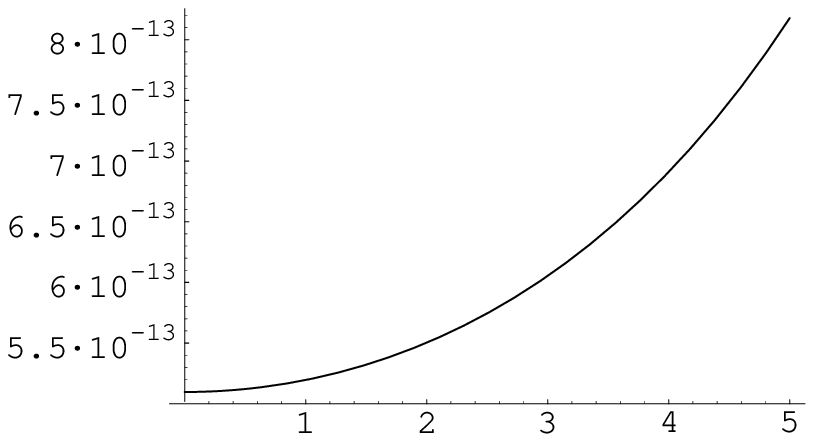}
  \caption{\small $\phi(r)\times r$ evaluated for $n=26$.}
\label{n1}
\end{figure}

\begin{figure}
\centering
\includegraphics[width=8.7cm]{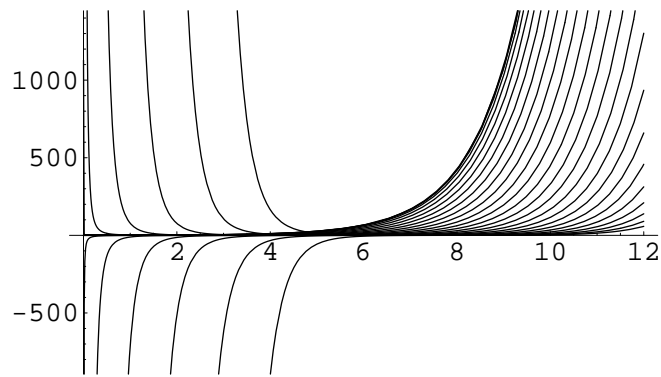}
  \caption{\small $\phi(r)\times r$ plotted for $1\leq n \leq 11$ and $n=26$ together.}
\label{n1}
\end{figure}

\end{document}